\documentclass[8.5pt]{article}
\usepackage[super,sort&compress,comma]{natbib} 
\usepackage{times,mathptmx}

\usepackage{amsmath} 
\usepackage{graphicx} 
\usepackage{epstopdf}
\usepackage[format=plain,justification=raggedright,singlelinecheck=false,font=small,labelfont=bf,labelsep=space]{caption} 
\usepackage{fancyhdr}
\usepackage[]{changes}

\begin{document}

\thispagestyle{plain}
\fancypagestyle{plain}{
\renewcommand{\headrulewidth}{1pt}}
\renewcommand{\thefootnote}{\fnsymbol{footnote}}
\renewcommand\footnoterule{\vspace*{1pt}%
\hrule width 3.4in height 0.4pt \vspace*{5pt}} 
\setcounter{secnumdepth}{5}

\makeatletter 
\def\subsubsection{\@startsection{subsubsection}{3}{10pt}{-1.25ex plus -1ex minus -.1ex}{0ex plus 0ex}{\normalsize\bf}} 
\def\paragraph{\@startsection{paragraph}{4}{10pt}{-1.25ex plus -1ex minus -.1ex}{0ex plus 0ex}{\normalsize\textit}} 
\renewcommand\@biblabel[1]{#1} 
\renewcommand\@makefntext[1]%
{\noindent\makebox[0pt][r]{\@thefnmark\,}#1}
\makeatother 
\renewcommand{\figurename}{\small{Fig.}~}

\fancyfoot{}
\fancyfoot[RO]{\footnotesize{\sffamily{1--\pageref{LastPage} ~\textbar \hspace{2pt}\thepage}}}
\fancyfoot[LE]{\footnotesize{\sffamily{\thepage~\textbar\hspace{3.45cm} 1--\pageref{LastPage}}}}
\fancyhead{}
\renewcommand{\headrulewidth}{1pt} 
\renewcommand{\footrulewidth}{1pt}
\setlength{\arrayrulewidth}{1pt}
\setlength{\columnsep}{6.5mm}
\setlength\bibsep{1pt}

\twocolumn[
 \begin{@twocolumnfalse}
\noindent\LARGE{\textbf{A Machine Learning Approach for Increased Throughput of Density Functional Theory Substitutional Alloy Studies}}
\vspace{0.6cm}

\noindent\large{\textbf{Alhassan S. Yasin\textit{$^{a}$} and Terence D. Musho,$^{\ast}$\textit{$^{a}$}}}\vspace{0.5cm}

\noindent\textit{\small{\textbf{Received Xth XXXXXXXXXX 20XX, Accepted Xth XXXXXXXXX 20XX\newline
First published on the web Xth XXXXXXXXXX 200X}}}

\vspace{0.6cm}

\textbf{Abstract}\\
\noindent \normalsize{In this study, a machine learning-based technique is developed to reduce the computational cost required to explore large design spaces of substitutional alloys. The first advancement is based on a neural network approach to predict the initial position of ions for both minority and majority ions before ion relaxation. The second advancement is to allow the neural network to predict the total energy for every possibility minority ion position and select the most stable configuration in the absence of relaxing each trial position. In this study, a bismuth oxide materials system, (Bi$_{x}$La$_{y}$Yb$_{z}$)$_2$ MoO$_6$, is used as the model system to demonstrate the developed method and potential computational speedup. Comparing a brute force method that requires the calculation of every possible minority concentration location and subsequent relaxation there was a 1.3x speedup if the neural network (NN) was allowed to predict the initial position prior to relaxation. This speedup is a result in an average decrease of 4 wall hour (64 cpu-hrs) reduction in relaxation for individual calculations. Implementation of the second advancement allowed the NN to predict the total energy for all possible trials prior to relaxation, resulting in a speedup of approximately 37x. Validation was done by comparing both position and energy between the NN to DFT calculations. A maximum vector mean squared error (MSE) of 1.6x10$^{-2}$ and a maximum energy MSE of 2.3x10$^{-7}$ was predicted. This method demonstrates a significant computational speedup, which has the potential for significant computational savings for larger compositional design spaces.}
\vspace{0.5cm}
 \end{@twocolumnfalse}
 ]

\footnotetext{\textit{$^{a}$~Department of Mechanical and Aerospace Engineering, West Virginia University, Morgantown, WV 26506-6106, USA. Fax: 304-293-6689; Tel: 304-293-3256; E-mail: tdmusho@mail.wvu.edu}}

\section{Introduction}
Advancements in first-principle material modeling techniques are finding applications for modern-day materials development that is required effective use of resources for new devices and structures. Modern materials require precise control of properties such as rapid phase response to external stimuli like pressure, light, magnetic field so that meaningful uses are possible in modern-day or can be expected in the near future. These modern materials (magnetic, ferroelectric, superconducting) are often multi-component systems such as but not limited to high-temperature superconductors, magnetic tunnel-junctions, and perovskite materials with complex magnetic structures~\cite{blugel05}. The underlying principle cohered with for these first-principle methods is that the parameters of the formulated theory are fixed by the basic assumptions and equations of quantum mechanics. 

During the past two decades, first-principle calculations based on density-functional theory (DFT)~\cite{kohn65} in the generalized gradient approximation (GGA)~\cite{dreizler13,jones89,ernzerhof96} or the local density approximation (LDA) unfolded as a successful approach to solve the electronic structure of matter~\cite{blugel05}. DFT is a widely used computational method that helps understand a wide range of material properties. The theory is able to reduce the many body Schr\"odinger equation to an effective single electron problem by relying on Hohenberg-Kohn theorem~\cite{hohenberg64} and Kohn-Sham method~\cite{kohn65}, thus making material property predictions computationally feasible~\cite{kanungo17}. The profound success of DFT for describing ground-state properties for vast classes of materials such as semiconductors, insulators, half metals, semimetals, transition metals, etc., at the nanostructure scale makes it one of the most used method for modern electronic structure analyses~\cite{blugel05}. Its noted that the goal of these calculations is to gain insight on a well defined model so that studies can find and predict trends that can better assist in developing different levels of understanding for any system in question~\cite{blugel05}.

Due to the extreme computational costs of most theoretical studies, limitations can and do arise when using approximation methods because accuracy is compromised in exchange for speed up time~\cite{smith17}. One of the most challenging aspects in modern theoretical calculations is to develop and apply an approximation method that expedites first-principle calculations speed up time without the loss of accuracy. Methods such as fragmenting the system~\cite{kitaura99}, construction of empirical potentials, corrections through statistical methods~\cite{li12}, linear scaling~\cite{ochsenfeld07}, or semi-empirical (SE) methods have been applied to attain accurate first-principles calculations that are still computationally effective~\cite{smith17,li12}. In the case of SE Tight Binding approximations to DFT, the time expense is reduced by treating the Hamiltonian elements as parameters adjusted for the desired properties of the system such as bandgap, effective mass, etc, also SE Tight Binding to DFT simplifies the Hamiltonian to nearest-neighbor interactions among atoms~\cite{hegde17}. The introduction of the exchange and correlation functionals allowed DFT to improve cost even further when compared to high-level first principle methods like MP2 and CI, with a very similar level of accuracy~\cite{li12}. However, even the fastest DFT techniques, such as O(N) approach that are often based on Local Orbitals, use up most of the computational time to iteratively formulate the Hamiltonian and solve self consistently for the ground state electron eigenstates~\cite{hegde17}. Thus relies heavily on the initial state of the system, the closer the final state of the system is to the initial state specified as the input, the fewer iterations spent formulating the Hamiltonian and solving self consistently the ground state. 

The reader should note that each molecule is unique and thus to explore different configurations, tailor or substitute various atoms into a structure, will only increase the computational cost. This is not to say one can avoid computational cost but in practicality greatly reduce it if the initial state of the system being explored is not just referenced based on some configuration but slightly refined based on the referenced configurations and based on what is being tailored or substituted. Thus exploring new configurations can be more computationally feasible, especially in a time-sensitive world where industrial applications rely heavily on a number of material systems such as geometries, boundary conditions need, and so on to be evaluated quickly and effectively to meet with the growing demand for production and application. Machine learning approaches have demonstrated viable solutions in attaining various forms of interacting and noninteracting atomistic potential by utilizing regression algorithms in resent years~\cite{smith17}. A variety of applications such as chemistry and physics have successfully applied machine learning methods to predict reaction pathways~\cite{jiang16}, formation energies~\cite{faber16}, excited energy states~\cite{hase16}, atomic forces and resonance chemical shifts, etc., to assist in searching and classifying material systems~\cite{smith17}. Also, computational material science has had several advancements made in applying various machine learning techniques such as predictions of DFT functionals~\cite{snyder12}, mapping of spacial atomic data for predicting total energies~\cite{behler07}, and computation of electronic properties~\cite{schutt14} in recent years. Thus machine learning express potential to predict molecular interactions with accuracy and reduction of computational cost.

\begin{figure}[h]
\includegraphics[width=1\columnwidth]{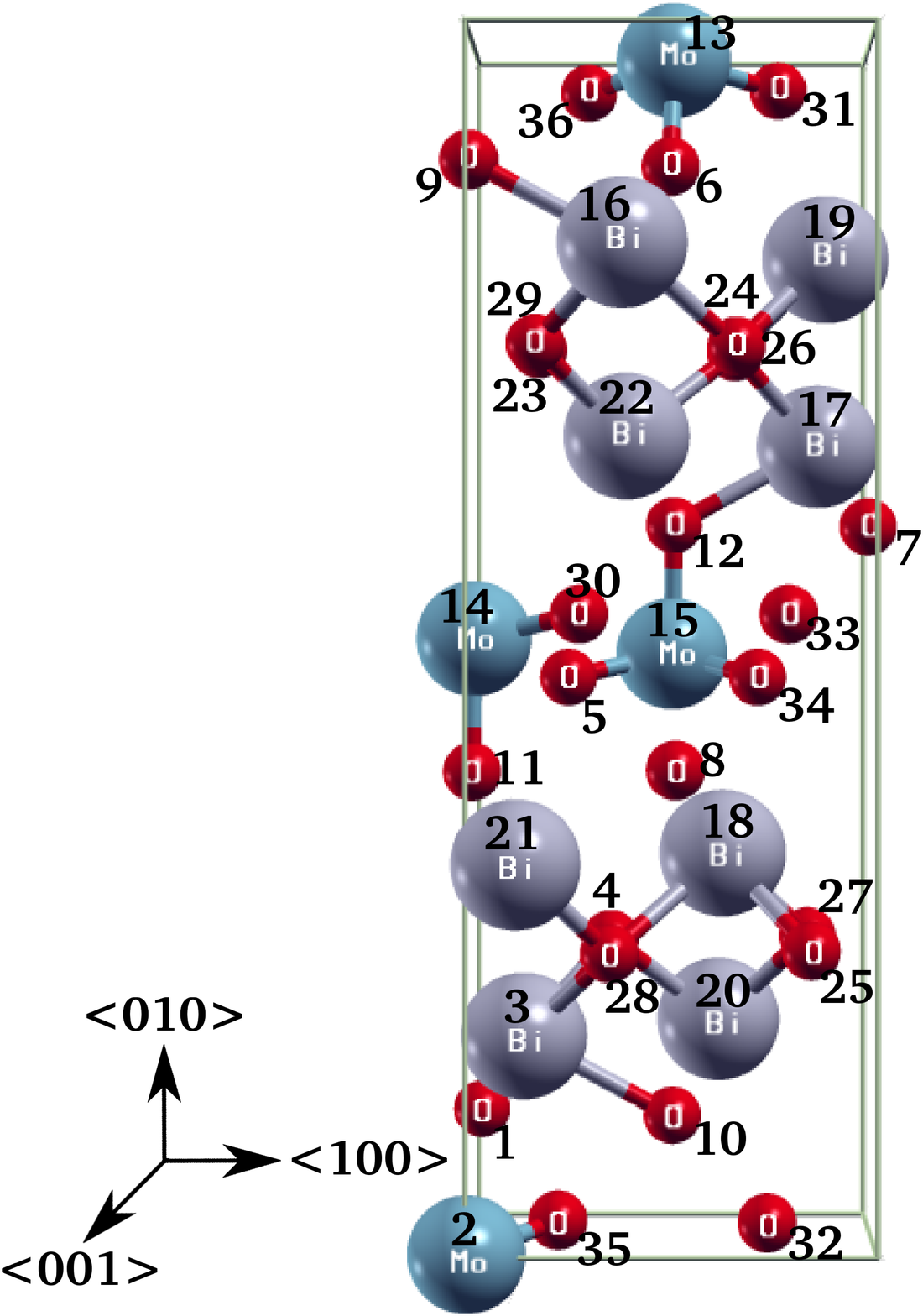}
\caption{Illustration of the Bi$_2$MoO$_6$ in the orthorhombic configuration that is used as the example structure system to test the machine learning approach for this study. The orthorhombic unit cell is made up of 24 O atoms 8 transition metal atoms (Bi) and 4 Mo atoms making a total of 36 atoms in a unit cell. La and Yb are subbed into the 8 metal atom positions (Bi) (atom position numbers 3, 16, 17, 18, 19, 20, 21, 22). These are the initial state of all DFT calculated structures and thus the numbers that correspond to each atom are the same for any configuration. All calculated DFT data is referencing the same atom position number to keep consistency, thus all attained calculations of position are for the same atom position number.}\hrule
\label{fig:uc}
\end{figure}

To address the issue of computational expense, this study proposes a machine learning-based method to predict positions of atoms (vectors that represent each atom) and the total ground state energy for each structure (unit cell). The predicted atomic positions that make up the unit cell and energy is then compared to that of the predicted DFT calculations. Note that this is predicting the fractional coordinates for each atom that makes up one primitive unit cell and the respected lattice constants. The machine learning model uses training sets based on DFT simulations done for various structures. This paper focuses on bismuth-based photocatalysts (Bi$_2$MoO$_6$) in the orthorhombic configuration (Figure~\ref{fig:uc}) and predictions made are validated with DFT calculations for configurations of (Bi$_{x}$M$_{y}$)$_2$MoO$_6$ where (M = La, Yb). However, this method is independent of a material system and depends only on referenced DFT simulations for a specific structure. The aim of this study is to predict fractional coordinates and the total ground state energy for a bismuth-based photocatalysts (Bi$_2$MoO$_6$) by taking already attained DFT calculations for (Bi$_{x}$La$_{y}$)$_2$MoO$_6$ configurations as the training set to a neural network and then predict the ground state energy and the fractional coordinate vectors with respect to the DFT calculations for (Bi$_{x}$Yb$_{y}$)$_2$MoO$_6$ in the orthorhombic configuration. By using the training set, a much better approximation of the initial state of the system can be attained which reduces the effective iterative steps taken in the DFT calculations to solve self consistently the ground state.

\section{Methodology}
Applying a neural network to any problem at hand requires input features (specified by user) to be mapped to some target output in some non-trivial way. In this paper, a supervised neural network (NN) is applied to a data set consisting of inputs (features that describe atomic classification in a unit cell) to be mapped on a desired output (expressing atomic positions in a unit cell and total ground state energy).

\begin{figure} [h]
\includegraphics[width=1\columnwidth]{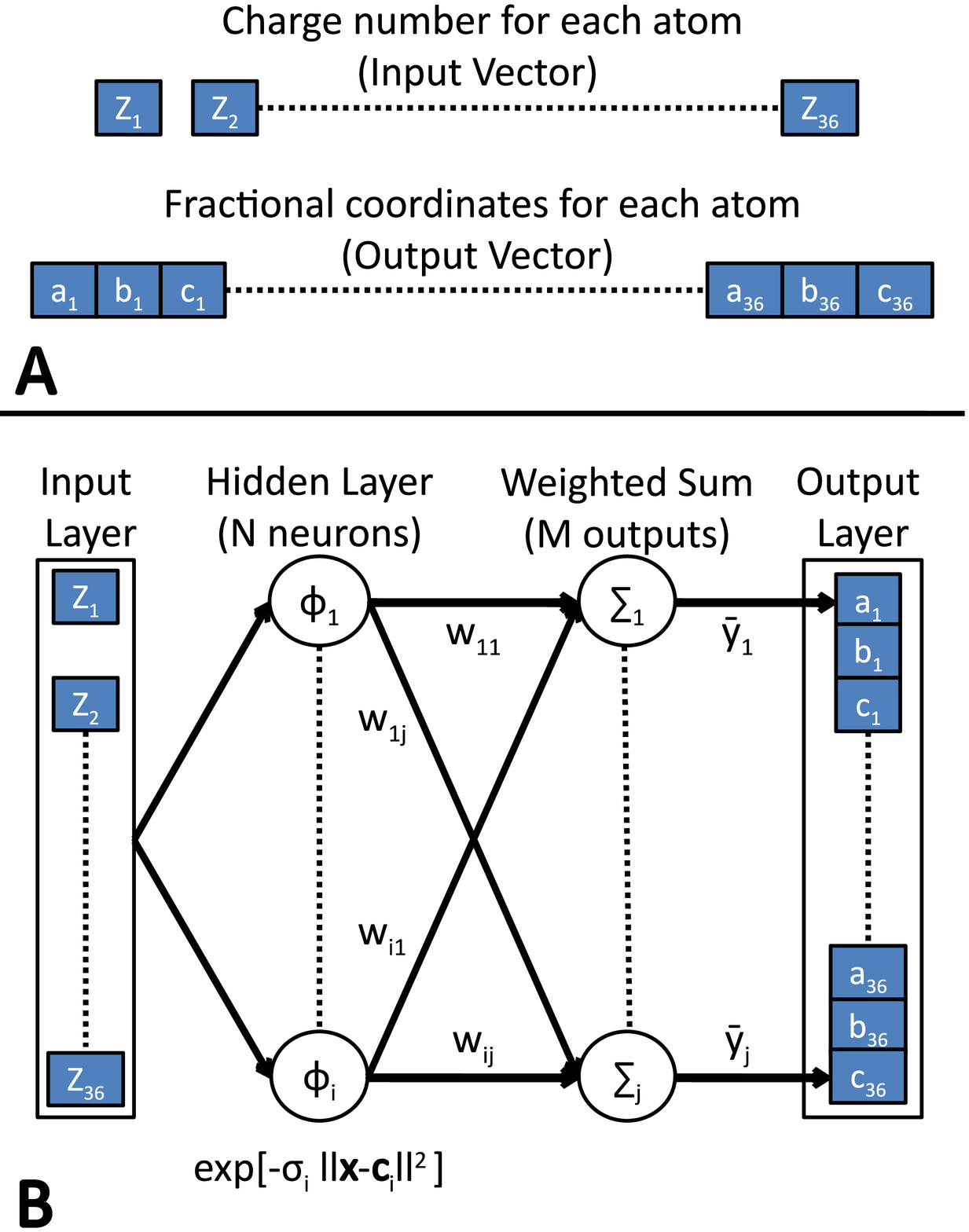}
\caption{Illustration A is the vector representation of input (charge number Z) and output (positions along the lattice vectors (\textbf{a}, \textbf{b}, \textbf{c}) in each unit cell). Illustration B is the artificial neural  network architecture implemented for this paper. It is noted that the numbers indicated in this figure (such as a$_{1}$ or Z$_{36}$) are referencing the same numbers initially started for each unit cell in Figure~\ref{fig:uc} to keep referencing atomic number positions in the NN consistent.}\hrule
\label{fig:nn_arc}
\end{figure}

\subsection{Model Design Space}
\label{sec:mds}
This study focused on the substitution of La and Yb into the bismuth-based structure (Bi$_2$MoO$_6$) as shown in Figure~\ref{fig:uc}. The main structure modification that was looked at is the substitution of La and Yb atoms for the Bi atoms in various combinations of the 8 possible positions occupied in the unit cell expressed in Figure~\ref{fig:uc} for atom position numbers 3, 16, 17, 18, 19, 20, 21, 22. Thus, all combinations of Bi and Yb, Bi and La are explored in an attempt to express a method for machine learning that can reduce the computational cost of exploring the substitution for a structured system. A usual method implemented for substitution is to calculate all possible combinations that the substitution can occur in the structure, which is very computationally intensive. To put this in perspective, for the structure system explored in this study, more then 300 unique configurations are evaluated, this is a classical permutation problem. For example, if to say we are looking at 1 Yb atom and all 7 Bi atoms implemented into the structure, Yb could potentially occupy any of the 8 positions in the unit cell, thus 8 unique combinations can be expressed for the  Yb 1/7 Bi ratio. Thus, this study aims to express a viable solution to the permutation problem so that not all structures are needed to be explored in a system, and if structures are explored, to express a better production of the initial state (initial configuration of the unit cell) of the system to reduce computational cost. The way proposed in this study is to use a machine learning technique to predict the fractional coordinates for all atoms that make up a unit cell (Figure~\ref{fig:uc}) and predict the total ground state energy for structure system which expresses a more viable solution to screen large design spaces. The model for the machine learning approach is DFT simulations of atomic structures, and also the validated results are compared to the DFT simulations.

\subsection{Input and Output Data Sets}
The data set consists of inputs in a set of atomic charges (atomic numbers) that make up a structure while the output of the network consists of fractional coordinates along lattice vectors (\textbf{a}, \textbf{b}, \textbf{c}) of each atom position that make up one unit cell and the corresponding total energy. The reference/training [(Bi$_{x}$La$_{y}$)$_2$MoO$_6$] and validation [(Bi$_{x}$Yb$_{y}$)$_2$MoO$_6$] data is attained by consistent DFT calculations. It is noted that all calculated DFT structures initially start in the same configuration, Figure~\ref{fig:uc}, which allows referencing input and output features consistent with the number that corresponds to each atom position for all unit cells as expressed by numbers in Figure~\ref{fig:uc}. The way the input and output features are expressed for the data set is by decomposing the input features (charges of atoms in the unit cell) and output features (atomic positions in the unit cell) as the vector representation expressed in Figure~\ref{fig:nn_arc}A. It is noted, those operations such as rotation, translation, and permutation to atoms in the data set (input and output) would ultimately change the ordering and positions of referenced atoms that make up the unit cell. Thus, it is very important that the input and output features formed for the neural network reflect the same corresponding number in the sequence of vector input and output given to the neural network for each unit cell. This is why the order in which the numbers expressed in Figure~\ref{fig:uc} is maintained when classifying input and output vectors for each unit cell represented as the vectors in Figure~\ref{fig:nn_arc}A. Thus allowing the network to progressively see how individual output feature changes as a consequence of input vectors.

To reduce the complexity of solving the atomic positions and gain understanding into fundamental and meaningful solutions for a given problem. This study imposes dimensionless quantities given to the neural network for data set interpretation, thus fractional coordinates are used for atomic positions. Understanding the energy as substitution occurs in the structure plays quite a central role in studies of chemical and biological systems. Thus, the ground state energy of the relaxed structure became another very important feature the neural network was trained to calculate. Keeping the same idea of dimensionless quantity in mind, this study proposes to express the energy as a ratio of the calculated sum of the individual energy that makes up the structure constituent parts divided by the total ground state energy, which is referred to as the energy ratio in this study.

\begin{equation}
Energy \: Ratio= \frac{\sum Individual \: Energy \: of \: Constituent \: Parts}{Total \: Energy \: of \: Structure}
\label{equ:nn_e}
\end{equation}
The ground state energy is the DFT calculated energy for the system and the individual energy of the constituents is determined from the DFT energy for individual atom (individual atoms in a big box). From this point, energy is expressed as an energy ratio. Unlike the output vector expressed for each input vector for fractional coordinates, the energy ratio is a scalar representation of energy for each input vector. Thus the total ground state energy for each system can be determined by using the sum of constituent atom energies divided by the energy ratio.

\subsection{Artificial Neural Network (ANN) Model}
This study aims at training a neural network with all configurations of (Bi$_{x}$La$_{y}$)$_2$MoO$_6$, inputs being the atomic number (number of protons) of the constitute atoms and the outputs being the fractional coordinates (position) for each atom in the unit cell, then to validate this with configurations of (Bi$_{x}$Yb$_{y}$)$_2$MoO$_6$. The general ANN model is a simple information processing unit with multiple inputs and outputs. A neurons within the architecture of the network is attaining inputs from other neurons or from the exterior through path modeling. The artificial neuron output is computed as the weighted sum of all inputs modified by an activation function. These weights are adjusted through the learning process of the ANN. The approach is to train the network with a given set of non-linear input to output data sets, in order to capture knowledge about the process. The method implemented in this study is that of an artificial neural network (ANN), where the network is comprised of three layers (Figure~\ref{fig:nn_arc}B), the input layer (IL), the hidden layer (HL) and the output layer (OL)~\cite{orr96}.

Each layer in the NN has a different task~\cite{yu06,howlett13}, the general architecture of the NN is illustrated in Figure~\ref{fig:nn_arc}B. From the IL to the HL of ANN the distance between the network input and hidden layer centers is calculated. From the HL to OL the weighted sum is computed for each neuron. Each neuron of the HL has a vector parameter called center, and the general expression of the network is given as~\cite{yu06,howlett13},

\begin{equation}
\bar{y}_{j}=\sum\limits_{i=1}^{N} w_{ij}\phi_{i},
\label{equ:1}
\end{equation}

where, $N$ is the number of neurons in the HL $(i \in \lbrace 1,2,..., N \rbrace)$, w$_{ij}$ are the weight of the $i^{th}$ neuron and $j^{th}$ output, $\bar{y}_{j}$ is the neural network's response to the $j^{th}$ output, and $\phi_{i}$ is the output for the $i^{th}$ neuron. The output of each neuron is referred to as the activation function which is taken as the Gaussian function defined as,

\begin{equation}
\phi_{i}=exp[-\sigma_{i} \Vert \textbf{x} - \textbf{c}_{i} \Vert^2],
\label{equ:2}
\end{equation}

where, $\sigma_{i}$ is the spread parameter of the $i^{th}$ neuron, \textbf{x} is the input data vector, $\textbf{c}_{i}$ is the center vector of the $i^{th}$ neuron, and $\Vert \textbf{x} - \textbf{c}_{i} \Vert^2$ is the Euclidean distance. Figure~\ref{fig:nn_arc}B expresses the architecture of the ANN used. Note that for the structure system this study is exploring the input vector ($\textbf{x}$) is a 1x36 vector (Z$_1$,...,Z$_{36}$), the vector is passed to the HL comprised of N neurons. At the HL each neuron calculates the Euclidean distance between its center vector and the input vector, then calculates the Gaussian function (activation function) with the spread parameter specified (Equation~\ref{equ:2}). Then the output layer calculates the weighted sum for each input multiplied by the neurons activation response (Equation~\ref{equ:1}) to that input vector to get the output of the network ($\bar{y}_{1},...,\bar{y}_{j}$), where M denotes the number of outputs $(j \in \lbrace 1,2,..., M \rbrace)$, being in this study the fractional coordinates.

The training part of the NN involves determining the number of neurons in the HL, and to attain the desired output for the network, the $w$, $\sigma$, and $c$ parameters can be adjusted and attained. The most common error reference response of the NN used is mean square error and sum square error to train the NN. Thus, for this study the error-based expression used (supervised learning) is defined as,

\begin{equation}
error(w,\sigma,c)=\sum\limits_{j=1}^{M} [y_{j} - \bar{y}_{j}]^2,
\label{equ:3}
\end{equation}

where, ${y}_{j}$ indicates the desired output, and $\bar{y}_{j}$ is the network output. Ultimately the training steps for the network entails the minimization of the error function. The training algorithm for the NN utilizes three approaches. The first is to use k-means clustering~\cite{hartigan79} for initially attaining the centers ($c$) for each neuron based on the input vectors. The second is to have the weights ($w$) updated based on the activation function for the training set by using pseudo-inverse~\cite{wettschereck92} of the activation function matrix used for all training sets. Lastly, using a gradient descent algorithm (GD)~\cite{karayiannis99} to progressively update the spread for each neuron ($\sigma$), thus ultimately minimizing the error function.

\subsection{Training Algorithm used for the Model} 
To attain the desired result for the neural network output to the training parameters, the training algorithm follows three steps. The reader should note that this section is only referencing the training set in order to teach the NN model, and not for the validation set. The first step used in the training algorithm is a k-means clustering~\cite{hartigan79} for input vectors of all training set to find the centers. The point of the k-means clustering is to partition all the input vectors of the training set into N clusters, N being the number of neurons specified, in which each input vector belongs to the nearest mean cluster, cluster is the center specified for the $i^{th}$ neuron. However, if the user wants to increases or decreases the number of neurons in the model a new k-means clustering has to be calculated to attain the new center for each neuron. 

Prior to the second step, the spread ($\sigma$) is initialized. Since the output of the HL multiplied by the weights is supposed to approximate the output training data as given in Equation~\ref{equ:1} or represented as follows,

\begin{equation}
\begin{bmatrix}
    \bar{y}^{(j)}_1       \\
       \vdots         \\
    \bar{y}^{(j)}_O      
\end{bmatrix} 
=
\begin{bmatrix}
    \phi_{1}(\textbf{x}_1,\sigma^{(j)}_1,c_1)  & ...&  \phi_{1}(\textbf{x}_O,\sigma^{(j)}_1,c_1)      \\
                     \vdots                      & \ddots&                  \vdots                           \\
    \phi_{N}(\textbf{x}_1,\sigma^{(j)}_N,c_N)  & ...&  \phi_{N}(\textbf{x}_O,\sigma^{(j)}_N,c_N)     
\end{bmatrix}^T
{\cdot}
\begin{bmatrix}
    w^{(j)}_1       \\
       \vdots           \\
    w^{(j)}_N,     
\end{bmatrix} 
\label{equ:4}
\end{equation}

where, $O$ is the number of training data sets, $N$ being the number of neurons, $c$ being the center that corespondents to each neuron, $j$ represents the output feature approximating for (such as a$_1$ in Figure~\ref{fig:nn_arc}B). For example $\bar{y}^{(1)}_O$ is the $a_1$ value for the $O$ vector input ($\textbf{x}_O$) whereas  $\bar{y}^{(2)}_O$ is the $b_1$ for the $O$ vector input, reference Figure~\ref{fig:nn_arc}B. A much simpler way to write Equation~\ref{equ:4} is as follows,

\begin{equation}
\bar{Y}^{(j)}=\Phi^{(j)T}{\cdot}W^{(j)}.
\label{equ:5}
\end{equation}

For the second step, the weights are computed by using the inverse of $\Phi^{(j)}$ if it is a square matrix, being the number of neurons is the same as the number of the training sets (N=O). Note that in this case the centers will just be the input vectors and the ANN will provide exact estimations for the training data, this is not recommended because the estimation will tend to overfit all desired inputs to the training data. However, if fewer neurons are used then will need to compute the pseudo-inverse~\cite{wettschereck92} of $\Phi^{(j)}$, which is done for most cases, thus the weights are expressed as,

\begin{equation}
 W^{(j)}=[\Phi^{(j)}]^{-1*} {\cdot} Y^{(j)}.
\label{equ:6}
\end{equation}

It is noted that Y$^{(j)}$ is the desired output for the training data and thus the weights are updated this way for each updated spread of the Gaussian function ($\sigma$). The final step in the training algorithm is updating the spread for each neuron ($\sigma$) for the next iteration of the training algorithm to further minimize the error in the NN output to the desired output of the training set. The method used is the gradient descent algorithm (GD)~\cite{karayiannis99} which is a first-order derivative-based optimization algorithm used for finding local minimas for a function. The method takes small steps proportional to the negative of the gradient for the error function at the current iteration to update the spread to the next iteration. The error function is defined as follows,

\begin{equation}
Error^{(j)}=\sum [Y^{(j)} - \bar{Y}^{(j)}]^2,
\label{equ:7}
\end{equation}

where $Y$ is the desired output of the training set, $\bar{Y}$ is the NN output, and $j$ is the output feature of the training set,  same $j$ represented in Equation~\ref{equ:4}-\ref{equ:8}. Thus, the GD~\cite{karayiannis99} algorithm is used to minimize the error and optimize the adjusting spread for each neuron by iteratively computing the partial derivative and updating the spreads ($\sigma$) in parallel. The spread takes the following form,

\begin{equation}
\sigma^{(j)}_{i+1}=\sigma^{(j)}_{i} - \eta \dfrac{\partial Error^{(j)}}{\partial \sigma^{(j)}_{i}},
\label{equ:8}
\end{equation}

where $j$ is the output feature of the training set, $i$ is the current iteration step, $\eta$ is a small step size (referred to as the learning rate). Thus a simple summary of training is as follows:

\begin{itemize}
\item[1)]  Select the number of neurons and initialize the spread parameter for each neuron.
\item[2)] For the training set do k-means~\cite{hartigan79} algorithm to attain the center for each neuron.
\item[3)] Compute the activation for each neuron and get the weights for each neuron by pseudo-inverse approach~\cite{wettschereck92}.
\item[4)] Check to see convergence, if not converged do a GD~\cite{karayiannis99} algorithm to update the spread for next iteration and repeat step 3. 
\end{itemize}

The convergence criteria specified for this study is that the training epoch are iterated until the NN training output data set stops improving in accuracy for 100 epochs. The optimization process implemented, is to carry out the training method 5 times using an order of magnitude smaller learning rate each time. It is noted that $w$, $\sigma$, and $c$ parameters rely on the number of neurons used for the network. Ultimately having more neurons than the needed amount causes the model to over fit the output to the training data set and increases the complexity of the network. Therefore the number of neurons used directly affects the performance of the network and has to be investigated based on desired result. 

Once training is done, approximating any set becomes relatively simple. First, the input for the desired set of atoms to be evaluated is processed through each neuron by computing the Euclidean distance to each center of each neuron specified in the training set. Second the neuron computes the activation function (Equation~\ref{equ:2}) which describes the relationship of input to the center of that neuron. Then lastly the approximated output for the desired input is computed as the sum of all neurons activation function multiplied by the corresponding weights attained in the training set, thus solving Equation~\ref{equ:5} for desired input and follows Figure~\ref{fig:nn_arc}B.

\subsection{DFT Computational Details}
Ground-state properties and total ground state energies are approximated for relaxed configurations of a bismuth oxide structure system (see Figure~\ref{fig:uc}), which assisted this study to implement a neural network that can analyze the relevant trends in data set by means of density functional theory approach~\cite{qe}. The DFT calculations used in this study implemented pseudo-wave function functional representation based on the Perdew-Burke-Ernzerhof (PBE) exchange-correlation function. The PBE exchange-correlation functional implemented at potentials with a cut-off energy wave function of 1496 eV (110 Ry) that was tested for good accurate and stable results for the studied unit cells of this paper. The various configurational combinations that the bismuth structure had and this study explored, benefited greatly in the reduction of computational expense due to the implementation of pseudized wave function. It is noted that all the DFT simulations for all configurations had the initial system to be that of the bismuth-based structure illustrated in Figure~\ref{fig:uc}. For all DFT simulations, a Monkhorst-Pack with a k-point mesh sampled at a 2x2x2 grid with 1/2,1/2,1/2 offset was implemented. The k-point grid was tested for both speed and accuracy. Because the unit cell is large the reciprocal space is smaller than what is typical for smaller real space unit cells. Thus 2 k-points in each direction was adequate. To account for Van der Waals interaction, a Van der Waals correction term~\cite{grimme06,barone09} is implemented in DFT simulations, however, this correction did introduce some empiricism into the calculations. It was determined that the Van der Waals interaction were necessary correct some over-binding of the metal/oxygen ions. A cut-off radius of 12 times that of the cut-off wave function is implemented, 1320 {\r{A}}, with the scaling parameter specified as 0.7 for the Van der Waals parameters. 

The unit cell for each configuration geometry was relaxed to relative total energy less than 1x10$^{-10}$ with an overall unit cell pressure less than 0.5 kBar by computationally solving the electronic density self-consistently. It is noted that DFT prediction of energy, bandgap, etc. is known to underpredict results because of the exchange-correlation terms and the over-analyticity of the used functionals. Thus, the calculated configurations and energies are used not as absolute but to express a relationship between input and output features (energy ratios) that the network model will be trained on to predict the trends for various configurations.
 
\section{Results and Discussion}
Understanding the characteristics that describe chemical and biological system behavior and attributes plays quite a role in guiding and predicting new material systems. The modeling tools used to simulate these quantum systems rely on first-principle calculations of electronic structures to predict interactions and attributes. This study proposes to use a NN to predict and refine the initial state of atomic configurations and ground state energy to reduce the computational expense needed for first-principle calculations based on DFT simulations. As mentioned in previous sections, we express a NN to a training set consisting of all configurations of (Bi$_{x}$La$_{y}$)$_2$MoO$_6$, and test (validate) results with configurations of (Bi$_{x}$Yb$_{y}$)$_2$MoO$_6$. Note the training data has no explicit knowledge of Yb in the training data. The NN will be trained on DFT results that were derived through a brute force method where every possible combination and position of the minority atom was calculated. The input to the NN is the vector input that describes the charges (proton number) that make up a structure and output being the desired feature. The first feature the NN is trained to approximate is the fractional coordinates (atomic positions) that make up a unit cell based on the bismuth structure (Figure~\ref{fig:uc}). The second feature the NN predicts is the energy ratio that describes the total ground state energy of a unit cell. 

\begin{figure}[!h]
\includegraphics[width=1.1\columnwidth,trim=1.7cm 0.2cm 2cm 1.5cm, clip=true]{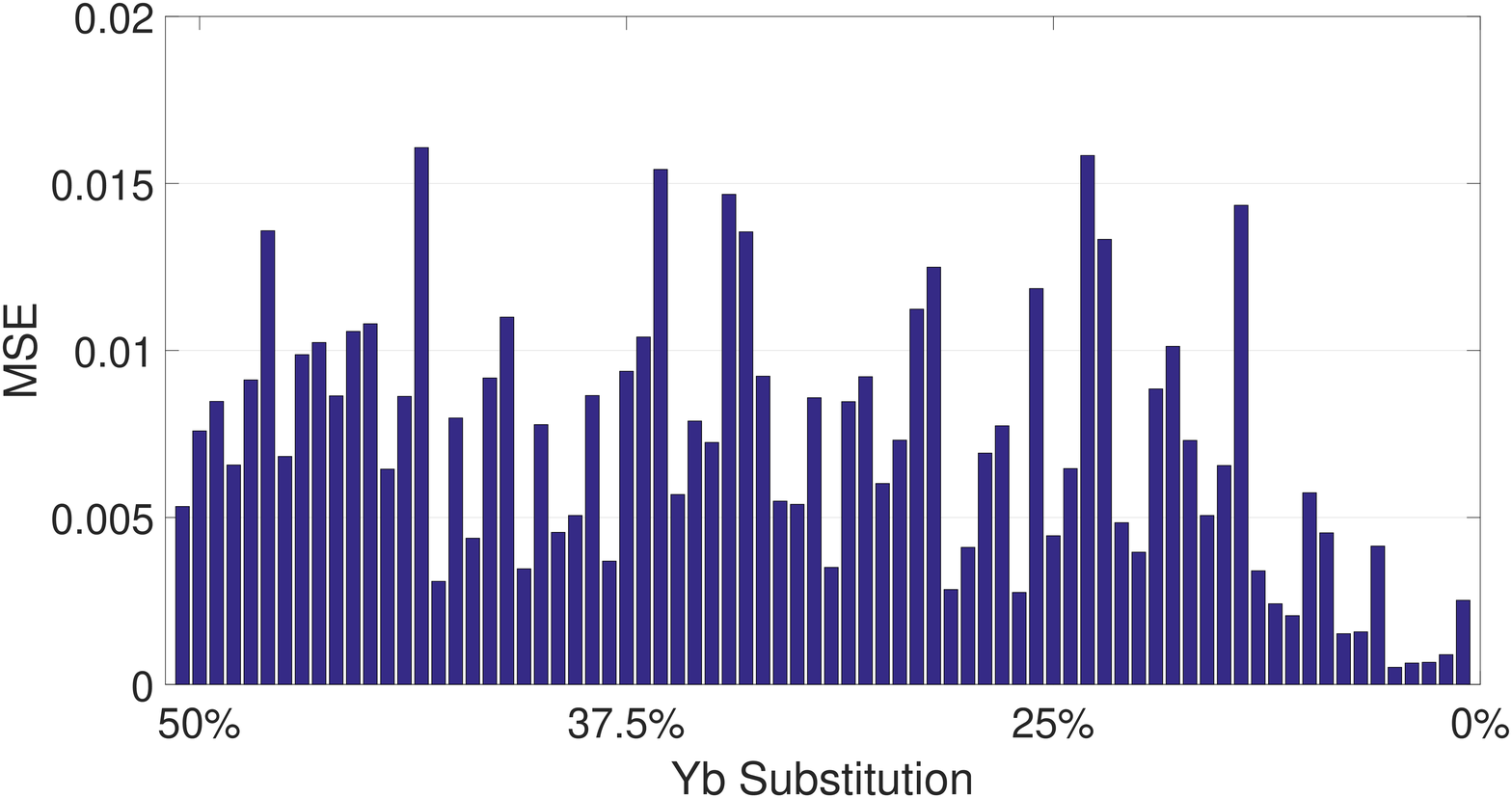} \put (-204,115) {\huge$\displaystyle*$}\\
\vspace{-1cm}
\begin{flushleft}
\textbf{\begin{LARGE}A\end{LARGE}}
\end{flushleft}
\includegraphics[width=1.1\columnwidth,trim=1.7cm 0.2cm 2cm 1.5cm, clip=true]{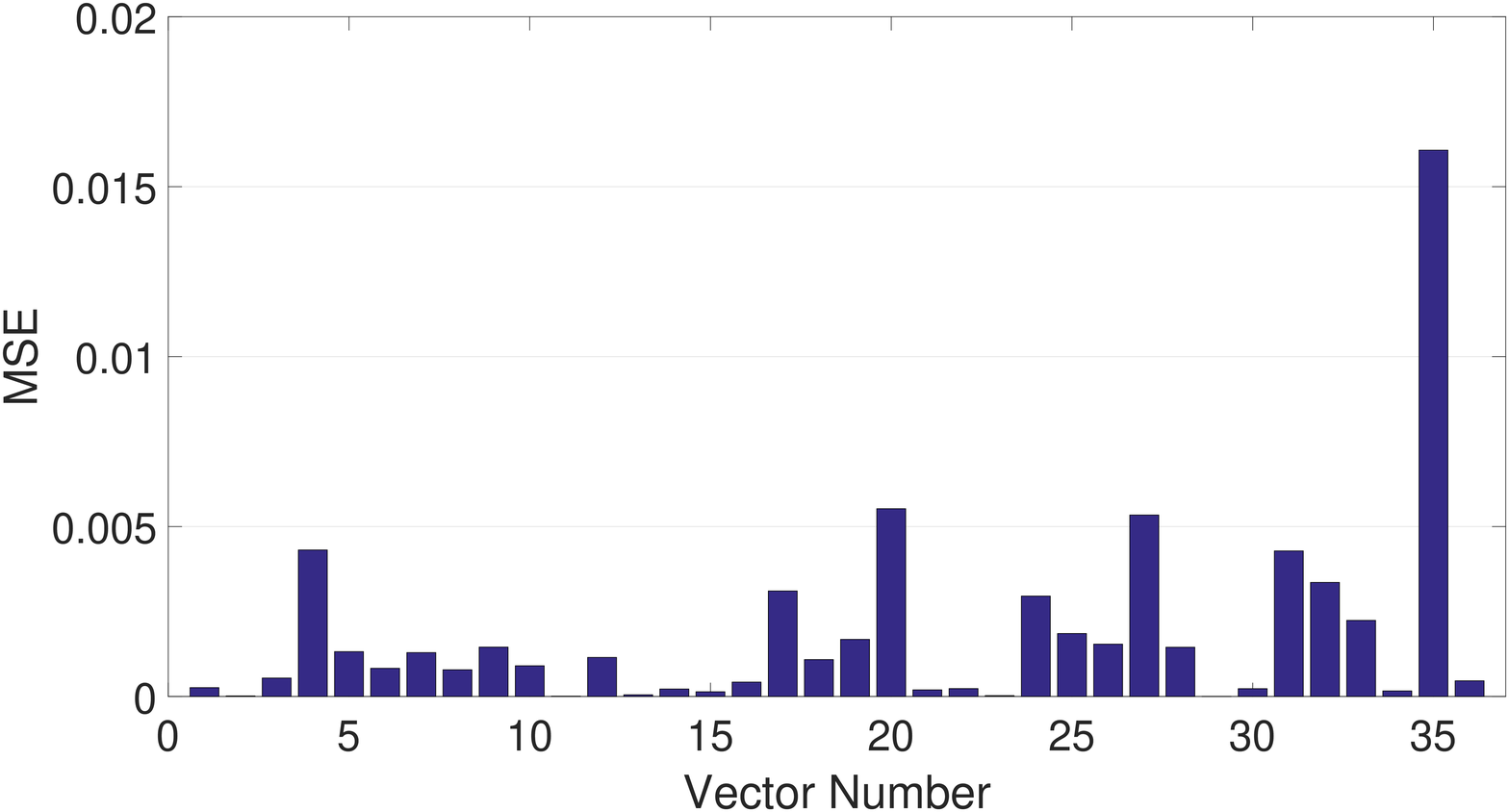} \\
\vspace{-1cm}
\begin{flushleft}
\textbf{\begin{LARGE}B\end{LARGE}}
\end{flushleft}
\caption{Plot A is the maximum vector mean squared error (MSE), Equation~\ref{equ:7}, for the structures with Yb substitution in the 8 atomic positions specified in Figure~\ref{fig:uc}, this is the error difference between NN and DFT. Plot B is showing the structure with the maximum error (worse case) difference between NN and DFT calculations for the 36 vectors and representing the '*' in plot A, the 36 vectors represent the 36 atomic positions in the unit cell expressed in Figure~\ref{fig:uc}. The vector errors shown in plot B correspond to 3 Yb atoms and 5 Bi atoms in the 8 positions as specified in Figure~\ref{fig:uc}, that is (Bi$_{5/8}$Yb$_{3/8}$)$_2$MoO$_6$. }\hrule
\label{fig:nn_responce}
\end{figure} 

\subsection{Approximating Atomic Positions}
\label{sec:ap}
Predicting atomic positions that make up a unit cell in a sense determines the computational time required to achieve convergence for that structure. In DFT, the total energy of any given system of interacting atoms and electrons is a function of the atomic positions that make up the structure and the electron density~\cite{blugel05}. The external potential used in DFT explicitly depends on the atomic position, which is changed by a small step to find an optimized atomic structure. Thus, the Hamiltonian and the wavefunctions used in DFT are also dependent on the atomic positions. The initial routine of DFT code is to solve the charge density self-consistently, which is solved iteratively by computing the potential terms in the Hamiltonian by initial guess of the input density and comparing that to the output density attained by using Kohn-Sham approach. This iterative step is considered converged when the self-consistent energy is within a specified accuracy. After this self-consistent calculation is done the atomic positions are moved by a small step, then a re-evaluation of the density and solving the self-consistent calculation is done till the problem is solved within some accuracy specified. Thus, the iterations required to solve this problem relies heavily on the atomic positions that make up the structure. It is quite simple to deduce that if a structured system has atomic positions initially close to that of the final atomic positions then the iterations requires to solve DFT calculations would be reduced. In practice, the initial atomic positions are specified by the structure system wanting to be tailored and or substituted, as in this study all DFT structure systems had initial atomic positions specified by Figure~\ref{fig:uc} which is the bismuth-based structure.

Figure~\ref{fig:nn_responce} illustrates the response of the neural network error for vector positions as Yb is substituted in the 8 atomic positions specified in Figure~\ref{fig:uc}. Its noted that for each unit cell there is a total of 36 vectors that correspond to the 36 atomic positions specified in Figure~\ref{fig:uc}. Thus, the maximum vector error is plotted for all cases as Yb is substituted, Figure~\ref{fig:nn_responce}A, and the configuration that yielded the maximum overall error, Figure~\ref{fig:nn_responce}B. Initial inspection of Figure~\ref{fig:nn_responce}A shows a random error of the NN response. However, that is not the case when looking carefully, there seems to be a trend, as more Yb substitution occurs the error progressively builds up. Which is to be expected, as lower concentration of Yb tends to be dominated by Bi, thus the training set which has seen high concentration of Bi will yield good results for higher Bi combination with Yb. Where as higher concentration of Yb will result in error build up due to the training set not consisting of any Yb substitution. Thus vectors of low concentration Yb will be evaluated more accurately, and based on the results the concentration of 1 Yb 7 Bi resulted in the lowest maximum overall vector MSE as 1.0x10$^{-3}$ Figure~\ref{fig:nn_responce}A. However, for the maximum error response overall of the NN as expressed by '*' in Figure~\ref{fig:nn_responce}A and Figure~\ref{fig:nn_responce}B, the highest maximum vector MSE is 1.6x10$^{-2}$ of 3 Yb 5 Bi. Its noted for both the maximum and minimum MSE for all Yb substitution, the vectors are compared to that of DFT, thus the error is reasonable compared to the computational expense DFT takes to attain the final structure. The only downfall is that the NN needs a good interpretation of training sets in order to accurately capture input and output features.

\begin{figure*}[!ht]
\includegraphics[width=1.0\columnwidth,trim=0cm 0cm 0cm 0cm, clip=true]{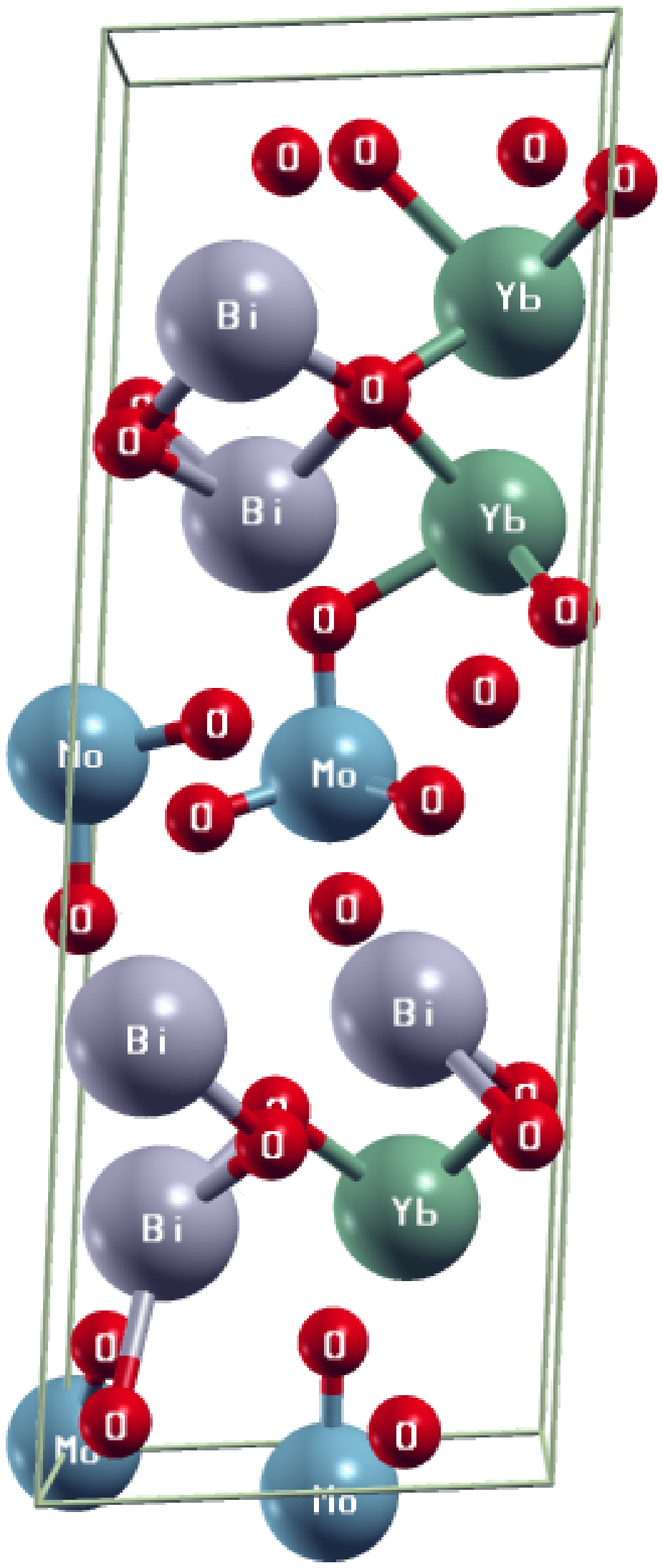}
\includegraphics[width=1.0\columnwidth,trim=0cm 0cm 0cm 0cm, clip=true]{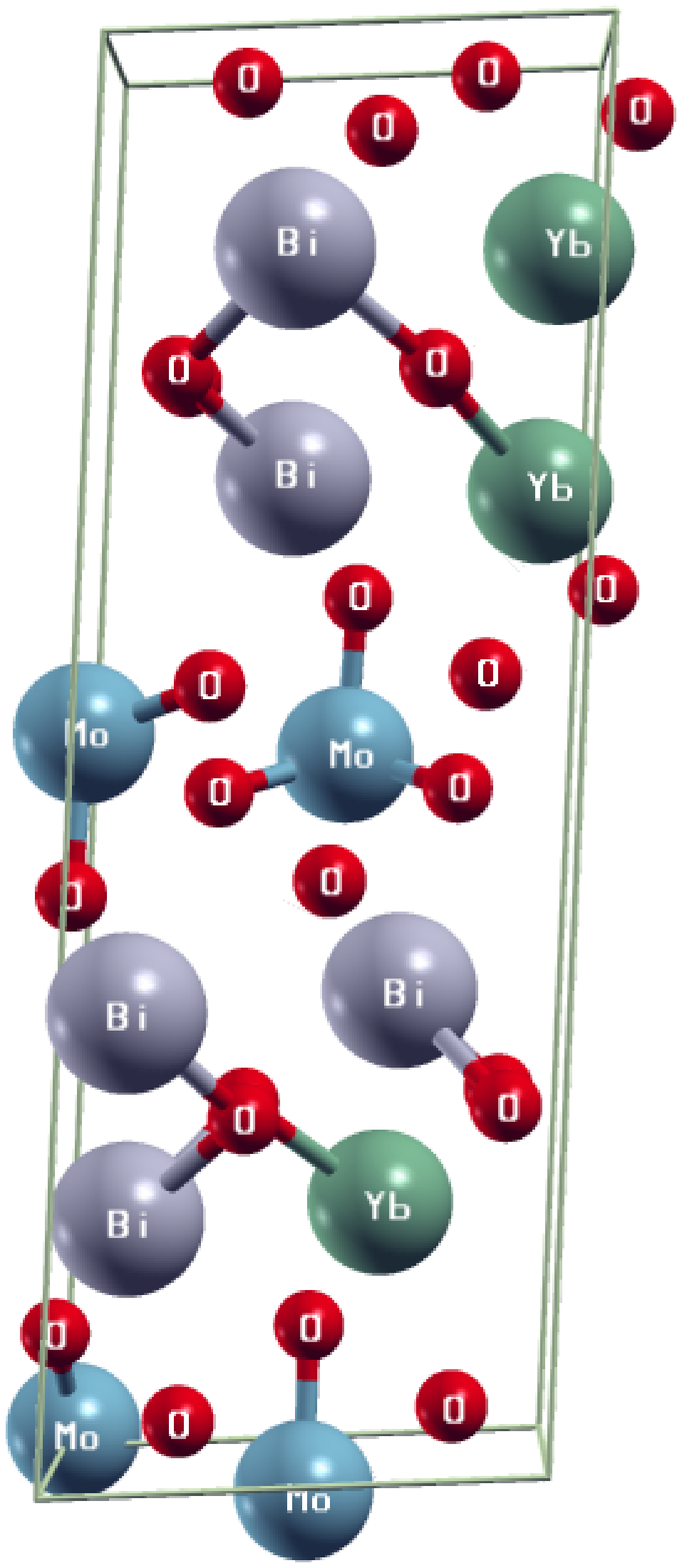} \\
\vspace{-1cm}
\begin{flushleft}
\textbf{\begin{LARGE}A\end{LARGE}}\hspace{87mm}\textbf{\begin{LARGE}B\end{LARGE}} 
\end{flushleft}
\caption{Considering the worst case, structure A represents the DFT attained optimized structure and B is the NN approximated structure for (Bi$_{5/8}$Yb$_{3/8}$)$_2$MoO$_6$. This configuration had the overall maximum error based on the NN approximations. Note that the individual vector MSE expressed in Figure~\ref{fig:nn_responce}B is comparing structure A (DFT) to B (NN). }\hrule
\label{fig:nn_dft}
\end{figure*}

Focusing on the worst cause, the overall error in the NN response, as in Figure~\ref{fig:nn_responce}B, atomic vector 35 is contributing to the maximum overall error for the (Bi$_{5/8}$Yb$_{3/8}$)$_2$MoO$_6$ structure. Furthermore, inspection of the atomic position that corresponds to vector 35 in Figure~\ref{fig:uc} reveals the location in the unit cell it occupies. This position is located at the boundary of the unit cell, at the boundary, the unit cell tends to be periodic in nature. Thus from an NN perspective, the training set had periodicity in the unit cell structures but was never told how this periodicity existed for these structures. The NN only saw how the atomic position would change from one input vector to the next, and so it was difficult for the NN to approximate the behavior of this structure at the boundary condition, a solution to this can be incorporating periodicity in the NN algorithm. However, for the intent of this study, it was more interesting to see if the NN could potentially predict the behavior at the boundary, and not by explicitly instructing the network to consider that behavior. After further inspection of the test set, most of the error was originating at the boundary conditions but this error was reasonably low. A visual interpretation of the error is expressed as vectors in Figure~\ref{fig:nn_responce}B for the (Bi$_{5/8}$Yb$_{3/8}$)$_2$MoO$_6$ structure.

It is very interesting to examine how the NN attempts to predict the structure configurations of Yb substitution. Comparing the optimized DFT structure, Figure~\ref{fig:nn_dft}A, and the NN predicted structure, Figure~\ref{fig:nn_dft}B, it is noted that the NN tends to distribute the atoms more evenly in the unit cell. This goes back to the training set, where configurations containing La expressed very similar interactions as configurations containing Bi, were structures expressed more even distribution of atoms in the unit cell. It is noted that all configurations of La, including pure La in the 8 positions proved stable from a DFT convergence perspective. However, not all Yb configurations were stable, and from DFT simulations anything more the 50\% would prove to not converge at all. Thus, this study tested (validate) Yb configurations to 50\% substitution in the 8 possible positions in the unit cell from DFT simulations. We also see when comparing A and B of Figure~\ref{fig:nn_dft} that the NN had done a better job at predicting the atomic positions that are within the unit cell. It is noted that atoms closer to the boundary tended to be more evenly distributed for the NN, such as the 4 oxygen sites located at the top of the unit cell in Figure~\ref{fig:nn_dft} A and B, this feature was expressed in configurations of the training set (La substitution) thus the network predicted a similar feature for most cases. However, with this error in mind, this study found it to be quite difficult for the NN to attaining exactly the atomic positions, this is quite an arduous task that requires very precise calculations of interactions in a structure system. Thus, the NN predicted structure is best suited as a initial screening method of approximate formation energies and determining the initial atomic positions that be subsequently used in the DFT model to more accurately capture boundary influences. In using the NN in this manner, the computational cost of using a random or full factorial calculation of the initial geometry is significantly reduced leading to decreased wall time for a given design space.

\begin{figure*}[!h]
\includegraphics[width=1.15\columnwidth,trim=2.5cm 0cm 4.2cm .2cm, clip=true]{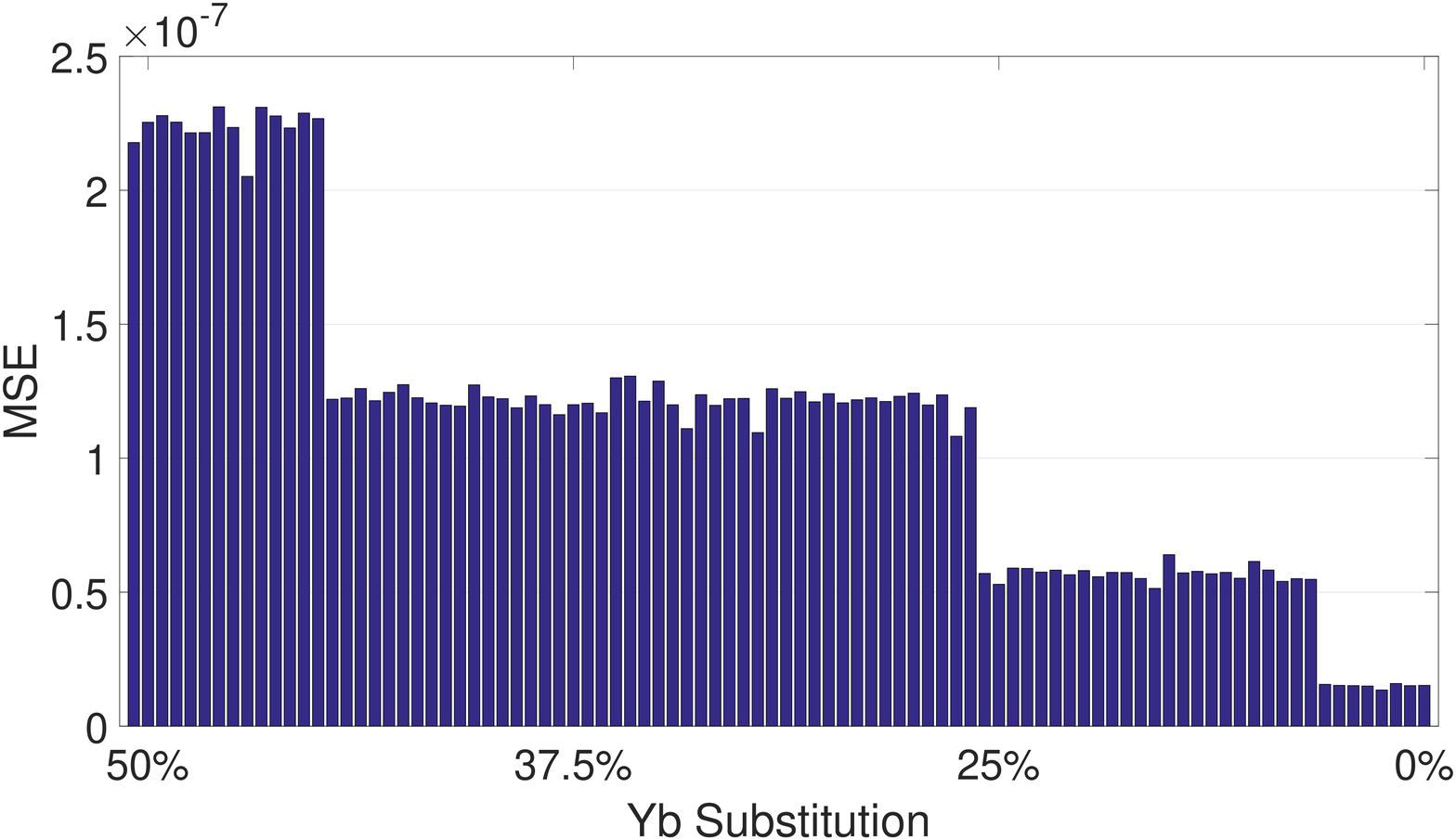}
\includegraphics[width=1\columnwidth,trim=0cm 0cm 0cm 0cm, clip=true]{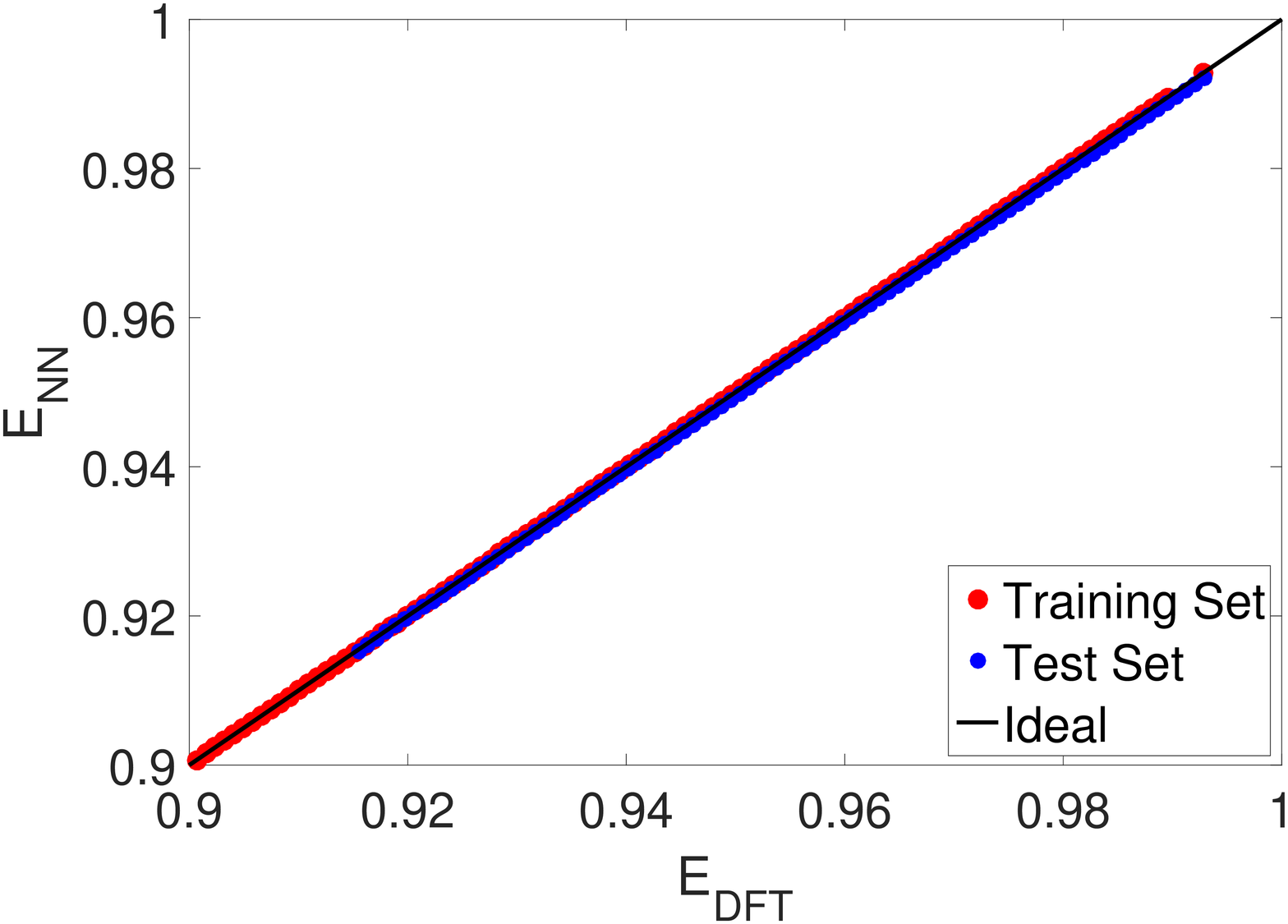} \\
\vspace{-1cm}
\begin{flushleft}
\textbf{\begin{LARGE}A\end{LARGE}}\hspace{100mm}\textbf{\begin{LARGE}B\end{LARGE}} 
\end{flushleft}
\caption{Plot A represents the energy ratio MSE for the structures contain Yb substituted in the 8 atomic positions. This error is comparing the NN predicted energy ratio to the DFT predicted energy ratio. Plot B expresses the predicted (E$_{NN}$) versus reference (E$_{DFT}$) energy ratio. The ideal line (x=y) is included to indict the quality of fit, when the predictions match the reference data perfectly, they will lie on the ideal line.}\hrule
\label{fig:energy}
\end{figure*}

\begin{table}[t]
\begin{center}
\begin{tabular}{ c |c  }
\hline
\multicolumn{2}{c}{(Bi$_{5/8}$Yb$_{3/8}$)$_2$MoO$_6$} \\ \hline
\hline
 Initial Positions Specified & Computational Time \\
 \hline
 Figure~\ref{fig:uc} & $\sim$30 hr \\
 Figure~\ref{fig:nn_dft}B & $\sim$26 hr \\
 \hline
\end{tabular}
\end{center}
\caption{The table expresses the computational time required for DFT simulation to find optimized structure, for the worst-case with maximum error, for (Bi$_{5/8}$Yb$_{3/8}$)$_2$MoO$_6$ configuration starting from initial atomic positions specified by Figure~\ref{fig:uc} and NN positions specified by Figure~\ref{fig:nn_dft}B. All parameters for DFT calculations are provided in the DFT Computational Details section, the only thing that is changed is the initial atomic positions.}
\label{table:improv}
\end{table}

The structure with the maximum overall error (worst case), (Bi$_{5/8}$Yb$_{3/8}$)$_2$ MoO$_6$, was re-evaluated using DFT simulation with the initial positions specified by the NN, Figure~\ref{fig:nn_dft}B, and evaluation of the computational time spent revels an overall improvement of 4 hours in wall time compared to the initial positions specified by Figure~\ref{fig:uc} for DFT simulation (Table~\ref{table:improv}). This improvement may seem non-trivial for a single structure, however when looking at the whole design space, this is quite an improvement. Overall the test set consisted of 100 configurations and each of the configurations had required an average of 30 hours of continuous running time for each simulation. Thus, the 4 hours would account for over 400 hour in reduction time in the overall study, if to presume each structure would be improved on average of 4 hours. However, this reduction is based on the time spend for the worst case with maximum error, and in practice most structures had much lower error, thus the improvement would be well over 400 hours. For instance, when testing the NN specified initial positions for the best case with the minimum error, being (Bi$_{7/8}$Yb$_{1/8}$)$_2$ MoO$_6$, the overall reduction in computational time is more then 9 hours, thus overall this study would account for way over 400 hours in computational wall time reduced. It is noted that most design studies of structure systems account for hundreds and even thousands of configurations, thus every possible method of improvement is explored, such as the expressed refinement of the initial system positions. Ultimately tailoring a structure system is a means to find a better structure that expresses very specific behaviors. In structure searching simulations, first principle calculations can be used to predict reaction steps associated with chemical reactions. An example is hydrogen evolution reactions (HER) or oxygen evolution reactions (OER), which require both minimization of the catalyst and the reactants. These reactions steps relies on the computation of the total energy for the structure system, which could be predicted by the NN, along with the initial position of reactants for subsequent DFT calculations. 

\subsection{Approximating Energy Ratio}
\label{sec:aer}
The total energy of the unit cell is ultimately affected by the atomic positions that make up the unit cell. Thus, it is possible to calculate the energy based on the predicted structure attained from the NN. However, this study intended for the NN to find relevant trends between input (proton charge) and the output features. With the error found based on the atomic vectors, it was reasoned that ultimately back calculating the energy would ultimately have a similar error. Ultimately, energy became an output to the NN not necessary associated with the NN predicted positions. It is noted that energy is given as an energy ratio to the NN, thus dimensionless and more manageable for this study (see Input and Output Data Sets Section). By having the NN to predict the energy and not back calculating it proved quite ideal as expressed by the error for Yb substitution in Figure~\ref{fig:energy}. Being able to predict the energy without the computational expense of a DFT prediction further decreases the computational demand of the permutation problem. This is because not all configurations will require DFT simulation to evaluate the energy. In some cases its more ideal to take the structure system with the lowest total energy, such as the example stated in the Model Design Space section. For instance, finding the best ratio (configuration with the lowest energy) of the Yb 1/7 Bi ratio requires DFT simulations of 8 unique configurations. Thus by using the NN to evaluate the energy one can get a sense of the relevant trend as Yb is configured in different positions in a unit cell.

Figure~\ref{fig:energy}A expresses the overall energy ratio MSE comparing the NN and DFT calculated energy for Yb substitution in the 8 positions. Its very interesting to see that the error expressed in Figure~\ref{fig:energy}A behaves like a step increase from low to high concentration of Yb. This is due to the evaluation of energy by the NN for similar configuration ratios like the 1/7 Yb to Bi. Thus, the energy for 1/7 Yb to Bi in one configuration is quite similar to a different configuration with the same ratio of 1/7 Yb to Bi. However, the error expressed for energy is of great accuracy, minimum error being 1.1x10$^{-8}$ MSE for low concentration of Yb (12.5\%) and maximum error of 2.3x10$^{-7}$ MSE for high concentration of Yb in the 8 positions (50\%). Figure~\ref{fig:energy}B expresses the quality of fit for the evaluated energy ratio expressed by the error in Figure~\ref{fig:energy}A, where comparison of the reference (DFT) versus predictions (NN) of energy for the training and tested set. The ideal line (x=y) in Figure~\ref{fig:energy}B indicates the quality for expressed predictions, thus predictions matching reference data (training set) would lie on the ideal line. It is quite reasonable to see that the energy attained from the NN is way within the energy computed by DFT simulations (ideal line in Figure~\ref{fig:energy}B), without the computational expense required to evaluate each structure configuration. In Figure~\ref{fig:energy}B, the red points indicate how well the NN responded to the training set, and expresses a visual interpretation of the NN performance for the training algorithm steps taken. The points indicated blue dots in the plot represent the tested set, being all the configurations of Yb up to 50\% substitution in the 8 positions. Ideally, the network should be within the line indicated as Ideal in Figure~\ref{fig:energy}B. This means that the energy calculated from DFT corresponds with the predicted NN energy ($E_{DFT} \approx E_{NN}$). Thus using the NN to predict energy proves quite ideal, and more so having the energy expressed as a ratio between the sum of the individual energy that makes up the structure constituent parts divided by the total ground state energy makes prediction straight forward.

\subsection{Computational Efficiency of the Proposed NN Approach}
Structure optimization studies involve huge design spaces, on the order of hundreds and even thousands of unique configurations. To evaluate the effective efficiency of the proposed NN approach, a control model must be compared and evaluated for the overall wall time required. The control model for this study is all the configurations that make up the training and tested set. The training set is comprised of 230 unique configurations of (Bi$_{x}$La$_{y}$)$_2$MoO$_6$ and the tested set are 100 configurations of (Bi$_{x}$Yb$_{y}$)$_2$MoO$_6$. Note that on average the wall time required for relaxation of each configuration took 30 hours of consistent DFT calculations on a 16 cores using an E5-2650 Ivy Bridge Intel processor (circa 2012). That is roughly 6900 wall hours (110,400 cpu-hrs) of continuous calculations for DFT simulation for the tested set and 3000 wall hours (48,000 cpu-hrs) for the training set. The total wall time required for the control model is mostly used to find the stable energy ratio configurations as expressed by Equation~\ref{equ:nn_e}, and in practicality there are vast amounts of structure analyses needed to be done for most structure based studies. However, to demonstrate the potential of the NN approach proposed by this study, the overall wall (and cpu) time is reduced by refining the initial atomic positions and predicting the energies prior to the execution of the DFT simulation for structure system. 

Because the initial position provided by the NN are close to the actual low energy position the computational expense of the DFT calculation are further reduced.) The overall wall time was shown to reduce by 4 wall hours (64 cpu-hrs) per configuration for the structure with the maximum error (worst case) and 9 wall hours (144 cpu-hrs) for the structure with the minimum error (best case). This is a result of the DFT relaxation calculation requiring less number of interations because the atoms are close to the equilibrium positions at the start of the calculation. As discussed previously, since most structures have MSE (position) errors way lower then the worst case, Figure~\ref{fig:nn_responce}A, the overall speed up is greater than what is calculated with the worst case speed up value. In this study, the overall improvement in the wall time exceeds 400 hours (6,400 cpu-hrs). The average wall time reduction for the tested set was calculated to be approximately 7 hours for each trail. That is calculating the average wall time for all the trails of the tested set, which ranged from 4 hours (worst case) to 9 hours (best case) in wall time reduction. Based on the calculated average wall (cpu) time reduced, an overall 700 wall hours of improvement for the tested set is realized. Thus reducing the 3000 wall hours of overall wall time to 2300 wall hours, which is a 1.3x speed up in computation for the tested set.

\begin{table*}[t]
\normalsize
\begin{center}
\begin{tabular}{ c |c |c |c  }
\hline
\multicolumn{4}{c}{(Bi$_{x}$Yb$_{y}$)$_2$MoO$_6$ (Tested Set Composition)} \\ \hline
\hline
 Method &	DFT Only & NN Refining Initial Atomic Positions & NN Energy Ratio Prediction \\ \hline
 Unique Configurations & 100	& 100	& 4 \\
 Overall Wall (CPU) Time& 3000 hr & 2300 hr & 120 hr \\
 Overall Speed Up & -- & 1.3x & 37x \\
 \hline
\end{tabular}
\end{center}
\caption[Overall wall time improvement made by the proposed NN approach]{This table expresses the computational wall (CPU) time required and improvement for three different approach to solve for the optimal configuration of (Bi$_{x}$Yb$_{y}$)$_2$MoO$_6$. The first method (DFT Only) is  the brute force purely DFT approach (baseline). The second method is to all the NN to predict the initial position of each configuration prior to the DFT calculation. The third method is to use the NN to predict the energy ratio for all configurations and down selecting the lowest energy configuration (4 configurations) and running those with DFT. Note that there are 100 unique configurations (trials) for the test set and on average each trial required continuous wall time of 30 hours.}\hrule
\label{table:ovarall_imp}
\vspace{-0.5cm}
\end{table*}

The second method of increasing performance is to reduce the number of configurations that have to be calculated by the DFT solver. This is done by using the NN predicted energy ratios for all possible configuration. Note for a single configuration (ex. (Bi$_{x}$Yb$_{y}$)$_2$MoO$_6$) there are 100 trials based on the combination of Bi and Yb that goes up to 50\% substitution. Therefore, only ratios of Yb occupying 1/8, 2/8, 3/8, 4/8 for the 8 possible positions in the structure system, as illustrated in Figure~\ref{fig:uc}. Thus by being able to predict the energies for the 100 trials prior to DFT simulation, a considerable amounts of configuration and wall time can be eliminated. Only configurations that represent the specific trial ratio will need to be evaluated. That is only 4 configurations of the 100 for the test set  need to calculated using DFT because there are only 4 ratios being evaluated for in the Yb case (1/8, 2/8, 3/8, 4/8). Thus, eliminating most of these trials greatly reduces the overall wall time of 3000 hours to 120 hours, which is a 37x speedup in the computation for the tested set. Table~\ref{table:ovarall_imp} is a outline of the overall wall time reduction by implementing the NN approach.

\subsection{Overall Feasibility}
Results in previous sections demonstrate not only the NN feasibility in learning the expressed system but NNs can provide accurate and meaningful features that greatly reduced DFT simulations computational time for individual run and for an collection of calculations. It would be ideal to demonstrate the proposed method to other DFT simulations of structure systems, for understanding the limitations and feasibility of the proposed method. The two features that the network was able to predict is the atomic positions and the total energy (formation energy). Further investigations regarding the best set of input features that could express the uniqueness of the structure may need to be looked at in future studies. However, the method expressed for this study, which is the inputs to the NN as proton charges of the constitute atoms proved to express reliable results. This was even more evident for the second output feature, which was the total energy.

Advantages of the expressed method versus alternate NN methods is the simplicity and ease of the algorithm implementation. The most arduous part was attaining the data input and output feature. Thus, the referenced and training sets were obtained by time consuming self-consistent calculations using DFT. The method expressed in this study links proton charge classification directly to structural information and total energy. While this approach is quite powerful and efficient for specific case studies, it does require having access to accurate DFT simulations for a subset of the design space. Thus limitations are based on the data set availability and having to train the NN on multiple configurations in order for a relevant trend to be characterized by the network. Thus this approach is ideal for large design studies. For instance the training algorithm consisted of approximately 230 training configurations of the (Bi$_{x}$La$_{y}$)$_2$MoO$_6$ and evaluated 100 configurations of (Bi$_{x}$Yb$_{y}$)$_2$MoO$_6$ for the test set. It is also noted that introducing any Yb configuration in the training part would prove to reduce the overall error because the network would see configurations similar to the trained set. However, this study focused on the feasibility of predicting and refining initial positions on configurations that are not trained explicitly by the NN. Thus this method is independent of material system and only depends on referencing DFT simulations for specific class of structure system like the bismuth bases structure (Bi$_2$MoO$_6$) explored in this study. Therefore, the proposed method is thus ideal for large design space based studies where structure systems could potentially take multiple substantiational attributes, thus making such studies reliant on computationally effective methods.

\section{Conclusion}
This study proposed a neural network technique based on machine learning for evaluating large design spaces to reduce the computational cost for DFT simulations. Predicting the optimized atomic structure and the total energy is ultimately the bases for most first principle calculations. Thus this study looks at the bismuth bases system (Bi$_2$MoO$_6$) as the example structure system and the design space consists of configurations in (Bi$_{x}$M$_{y}$)$_2$ MoO$_6$ where (M = La, Yb). We have demonstrated that this method can provide accurate predictions that will ultimately reduce the computational cost required to explore the structure unique configurations by applying one of the two methods. The first method to reduce the computational cost is by having the neural network better predict and refine the initial structure system based on a given set of training data. This way the DFT simulations will have a more refined method to initialize the unit cell being evaluated, which will allow less iterative steps to be taken in evaluating the optimized configuration for that specific structure. The second method proposed, exposes the network to a training data set for a given structure system, in order for the total energy of the optimized configuration to be predicted. This way the total energy of any configuration that makes up a given structure system can be predicted accurately based on already attained training data set for that structure system. 

The training set used for this study are configurations of (Bi$_{x}$La$_{y}$)$_2$MoO$_6$, and the tested set used are of (Bi$_{x}$Yb$_{y}$)$_2$MoO$_6$ configurations. Both the training and the test sets are evaluated with DFT simulations in order to quantify the error response from the DFT simulations to the neural network predictions. Ultimately for the bismuth-based system explored for this study, an overall maximum vector means squared error of 1.6x10$^{-2}$ was predicted by the neural network for structure configuration of (Bi$_{5/8}$Yb$_{3/8}$)$_2$MoO$_6$ being the worst case. This NN predicted structure was subsequently used as the initial position in a DFT calculation, realizing a 4 hour wall time (64 cpu-hrs) reduction in DFT computational time. For the tested set, this overall reduction would account for a cumulative time of 400 wall hours (6,400 cpu-hrs). Moreover, this study went a step further to all the NN to predict the lowest energy configuration. Thus reducing the number of DFT trial simulation from 100 per alloy composition to 4. This resulted in a realized speed up of 37 times or for the material system of interest in study a savings of 2880 wall hours (46,080 cpu-hrs). Ultimately the machine learning technique developed and implemented for this study provides a promising starting point for high-throughput electronic structure predictions for DFT simulations in large design space-based studies.

\section{Acknowledgments}
This research was conducted in support by the Super Computing System (Spruce Knob) at WVU, which is funded in part by the National Science Foundation EPSCoR Research Infrastructure Improvement Cooperative Agreement \#1003907, the state of West Virginia (WVEPSCoR via the Higher Education Policy Commission) and WVU. Author A.Y. would also like to acknowledge the partial support by U.S. Department of Energy ARPA-E Grant DE-AR0000698.

\section{Competing Interest}
The authors declare no competing interests.

\section{Data Availability}
The data that support the findings of this study are available from the corresponding author upon reasonable request.

\section{Author Contribution}
Author A. Y. has contributed to the article by conducting the DFT simulations, generating plots, contributed to writing of the article, and analysis of data. Author T. M. has contributed to writing the article and analysis of data.

\footnotesize{
\bibliography{journal} 
\bibliographystyle{plain} 
}


\end{document}